\documentstyle[psfig,prl,aps]{revtex}
\begin{document}
\begin{center}
{\Large\bf Ordering in the dilute
weakly-anisotropic antiferromagnet $Mn_{0.35}Zn_{0.65}F_2$}

\vspace{0.25in}

{\large F. C. Montenegro,$^1$ D. P. Belanger,$^2$ Z. Slani\v{c},$^2$
and J. A. Fernandez-Baca$^3$}\\
{\small $^1$Departamento de Fisica, Universidade Federal de Pernambuco,
50670-901 Recife PE, Brasil}\\
{\small $^2$Department of Physics, University of California,
Santa Cruz, CA 95064 USA}\\
{\small $^3$Solid State Division, Oak Ridge National Laboratory,
Oak Ridge, TN 37831-6393 USA}
\end{center}
\begin{abstract}
The highly diluted antiferromagnet $Mn_{0.35}Zn_{0.65}F_2$ has been
investigated by neutron scattering in zero field.
The Bragg peaks observed below the N\'{e}el temperature
($T_N \approx 10.9$~K) indicate stable antiferromagnetic long-range ordering
at low temperature.  The critical behavior is governed by random-exchange
Ising model critical exponents ($\nu \approx 0.69$ and $\gamma \approx
 1.31$),
as reported for $Mn_{x}Zn_{1-x}F_2$ with higher $x$ and for the
 isostructural
compound $Fe_{x}Zn_{1-x}F_2$.  However, in addition to the Bragg peaks,
unusual scattering behavior appears for
$|q|>0$ below a glassy temperature $T_g \approx 7.0$~K.  The glassy region
$T<T_g$ corresponds to that of noticeable frequency dependence in
earlier zero-field ac susceptibility measurements on this sample.
These results indicate that long-range order coexists with short-range
nonequilibrium clusters in this highly diluted magnet.

\end{abstract}
\vspace{0.4in}

Diluted uniaxial antiferromagnets have been extensively studied
as physical realizations of theoretical models of random
 magnetism\cite{by91},
including those pertaining to percolation phenomena\cite{e80,sa94}.
For three dimensions ($d=3$), two of the most extensively studied
examples are the rutile compounds $Mn_{x}Zn_{1-x}F_2$ and
 $Fe_{x}Zn_{1-x}F_2$.
These two systems differ effectively only in the strength and nature
of the anisotropy, providing a unique opportunity to explore the role of
anisotropy in the ordering of dilute magnets at low temperature.
In $Mn_{x}Zn_{1-x}F_2$ the anisotropy is dipolar in
origin\cite{nld69}.  In $Fe_{x}Zn_{1-x}F_2$ the anisotropy is
an order of magnitude greater for $x=1$ because of the additional crystal field
contribution\cite{hrg70}.  In many experiments with the magnetic
concentration, $x$, well above
the percolation threshold concentration $x_p=0.245$\cite{e80}, the
 behaviors
for $H=0$ are qualitatively similar for $Mn_{x}Zn_{1-x}F_2$ and
$Fe_{x}Zn_{1-x}F_2$.  Antiferromagnetic (AF) long-range order (LRO) at low
temperatures and characteristic random-exchange Ising critical behavior  
have been observed in the $Fe_{x}Zn_{1-x}F_2$ compounds for
$x \ge 0.31$\cite{mkjhb91}.  Similar random-exchange Ising model (REIM)
behavior is found in the
$Mn_{x}Zn_{1-x}F_2$ system for $x>0.4$\cite{rkj88}.

For small fields applied
parallel to the uniaxial direction and reasonably small magnetic dilution, the
diluted antiferromagnet in a field (DAFF) is expected\cite{fa79,c84} to show
critical
behavior belonging to the same universality class as the random-field Ising
model (RFIM)
for the ferromagnet, the latter being the model most used in
simulations\cite{r95}.
Indeed, for all measured samples of both systems for which the REIM
character is found at
$H=0$, the application of a small field parallel to the
easy axis generates critical behavior compatible with the
predicted\cite{fa79}
REIM to RFIM crossover scaling. In spite of the evidence supporting the
DAFF as a realization of the RFIM, some
non-equilibrium features
inherent to DAFF compounds and also the newly explored field
limits\cite{mltl98,mltl99,rfm99}
of the weak RFIM problem in $d=3$ make the nature of the phase transition at
$T_{c}(H)$ still a matter of considerable controversy\cite{b98,bfhhrt95,bkm96}.

Under strong random fields
(corresponding to large H) and also close to the percolation threshold, the
phase diagrams
of DAFF's have proven to be much more complicated than originally anticipated.
For large $H$, AF LRO is
predicted to become unstable\cite{rgsl85}. The generation of strong random
fields
induces\cite{ab87,nu91} a glassy phase in the upper part of the ($H$,$T$)
phase diagram
of $d=3$ Ising DAFF's. The equilibrium boundary,
\begin{equation}
T_{eq}(H)= T_{N} -bH^{2} -C_{eq}H^{2/\phi},
\end{equation}
above which hysteresis is not observed, has a convex shape at
high $H$ ($\phi > 2$), instead of the concave ($\phi=1.4$) curvature seen at
low field
(where REIM to RFIM crossover occurs). This change of curvature in $T_{eq}$
vs. $H$
was first
observed\cite{mlcr90} by magnetization measurements in
$Fe_{0.31}Zn_{0.69}F_{2}$.
Faraday rotation\cite{mkjhb91} and neutron scattering\cite{bmmkj91}
experiments on a sample with the same $x$, confirmed the REIM to RFIM
crossover scaling
at low $H$ and the lack of stability of the AF LRO at large $H$, giving
way to a random-field
induced glassy phase in this highly diluted compound. Recent magnetization
measurements
indicated that similar structure in the phase diagram exists at very high
fields for samples
with higher values of $x$\cite{mltl98,mltl99}. At a still higher
concentration the low temperature
hysteresis observed for $x<0.8$ is absent\cite{sbf97}.

The magnetic features observed at large $H$
in samples of $Fe_{x}Zn_{1-x}F_{2}$ in the concentration range $0.3<x<0.6$
contrast with the
behavior in weakly anisotropic system $Mn_{x}Zn_{1-x}F_{2}$ for intermediate
$x$,
where a strong $H$ induces a spin-flop
phase\cite{mjmmr92}. This distinct behavior may be solely a consequence of
the stronger Ising
character of the former system. In the strong dilution regime ($x \approx x_{p}$),
a number of magnetic features lead us to distinguish between these two
systems, as well.
For $x \leq 0.27$, no long range order\cite{by93} is observed in
$Fe_{x}Zn_{1-x}F_{2}$.
Typical spin-glass behavior was found\cite{mrc88} in a sample with
$x=0.25$, although recent works\cite{jdnb97,si98,jnm99} suggest
non-critical dynamics for $x$
close to $x_{p}$ in this system. Close to the percolation threshold even a
minute exchange frustration is a suitable mechanism\cite{syp79} for the
spin-glass phase in
$Fe_{x}Zn_{1-x}F_{2}$, as supported by local
mean-field simulations\cite{rcm95}. For Ising systems, it is
also expected that the dynamics even
at zero field should be extremely slow\cite{h85}. In $Mn_{x}Zn_{1-x}F_{2}$,
ac susceptibility
measurements indicates a spin-glass clustering at low temperatures for
samples with $Mn$
concentrations $0.2<x<0.35$\cite{bp81,mrjmm95}. Earlier neutron scattering
studies\cite{csbsg80}
suggest, however, that at $H=0$ the termination of the line of the
AF-paramagnetic (P)
continuous phase transition occurs at $T=0$ at $x=x_{p}$ in stark contrast to
the behavior of $Fe_{0.25}Zn_{0.75}F_2$.  In light of this contrast, the
influence of the frozen spin-glass
clusters on the stability of the AF LRO for $x$ close to $x_{p}$  in
$Mn_{x}Zn_{1-x}F_{2}$ is an important question that motivated the present
work. The dipolar anisotropy
of this weakly anisotropic system is expected to become random in strength
and direction as $x$
decreases, in contrast to the $x$-independent single-ion anisotropy of
$Fe_{x}Zn_{1-x}F_{2}$. 
In the case of $Mn_{x}Zn_{1-x}F_{2}$ under strong dilution, the application of
the results from numerical simulations\cite{rcm95} applied to
Ising systems is of course
not warranted.  Any differences
observed in this
system and the $Fe_{0.31}Zn_{0.69}F_{2}$ must certainly reflect the difference
in anisotropy and
this may give a window to the understanding of the general phase diagrams for
dilute anisotropic
antiferromagnets in applied fields.

In this study we performed zero-field neutron scattering experiments in
$Mn_{0.35}Zn_{0.65}F_{2}$
to verify the existence of a stable long-range ordered antiferromagnetic
phase below a
critical temperature $T_N \approx 10.9$~K, where REIM critical exponents $\nu
\approx 0.69$ and
$\gamma \approx 1.31$ govern the behavior and to investigate the
dynamic features of the system at low temperature. An unusual
scattering behavior
appears, for $|q|>0$, below $T_g \approx 7.0$~K, corresponding to the
region where earlier
ac susceptibility studies\cite{mrjmm95} indicated a noticeable frequency
dependence in
the real part of the susceptibility, in the absence of external field. The
results indicate
that long-range order coexists with non-equilibrium clusters in this highly
diluted system.

The neutron scattering experiments were performed at the Oak Ridge National
Laboratory using the HB2 spectrometer in a two-axis configuration
at the High Flux Isotope Reactor.  We used the (002) reflection of
pyrolitic graphite to monochromate the beam at $14.7$~meV.  The
collimation was 60 minutes of arc before the monochromator, 40 between the
monochromator and sample, and 40 after the sample.
A pyrolitic graphite filter reduced higher-energy neutron contamination.
The c-axis of the crystal was vertical and parallel to the applied field.
A small mosaic was observed from the Bragg peak scans at
low temperature, with roughly a half-width of $0.2$ degrees of arc
or $0.0035$ reciprocal lattice units (rlu).  The mosaic was incorporated
into the resolution correction by numerically convoluting the
measured resolution functions, including the mosaic, with
the line shapes used in the data fits\cite{by87}.
Most of the scans taken were (1 q 0) transverse scans.
For simplicity, the line shapes used in the fits to the data
are of the mean-field form
\begin{equation}
S(q)=\frac{A}{q^2+\kappa^2}+M_s^2\delta (q) \quad ,
\end{equation}
where $\kappa = 1/ \xi $ is the inverse fluctuation correlation length
and $M_s$ is the Bragg scattering from the long-range staggered
magnetization.  The critical power-law behaviors are expected to be
$\kappa = \kappa _o ^{\pm} |t|^{\nu}$, $\chi = A\kappa ^{-2} =\chi _o ^{\pm} |t|^{-\gamma}$
and $M_s = M _o |t|^{\beta}$, where $M_o$ is non-zero only for $t<0$.
The exponents $\nu$, $\gamma$, and $\beta$ and amplitude ratios
${\kappa_o} ^+/{\kappa_o} ^-$ and ${\chi_o}^+/{\chi_o}^-$ are universal
parameters characterizing the random-exchange Ising model.

The sample was wrapped in aluminum foil and mounted on an
aluminum cold finger.  A calibrated carbon resistor was
used to measure the temperature.

Transverse scans, taken after quenching to low temperatures
($T=5K$) and subsequently heating the sample, are shown in Fig.\ 1.
For clarity, the data for the range $|q|<0.008$~rlu, which spans the
Bragg scattering
component, are not shown.
For the most part, the scans are quite consistent with what is expected
for a phase transition occurring near $T=11K$.  However, a
most unusual feature of the line shapes is evident in the data
at the lowest temperature, $T=5$~K.  The broad line shape indicates
a great deal of short range order present upon quenching.  The short-range
order is evident for the scans with $T<7$~K.  The scan at $T=6$~K shows
striking asymmetry as shown in Fig.\ 2.
Since the scans were taken with increasing $q$ from $-0.19$
to $0.19$~rlu and each measurement took about $35$ seconds, the asymmetry
is an indication that the short-range order is rapidly decreasing
with time, i.e. the system is equilibrating.  The slow relaxation for
$T<7$~K corresponds very well to the large frequency dependence
observed using ac susceptibility in the same sample\cite{mrjmm95}
for $T<7$~K.

A transition to antiferromagnetic long-range order is indicated
by the presence of a resolution limited Bragg scattering
peak which decreases sharply as $T$ approaches $T_N \approx 11$~K.
As $T \rightarrow T_N$ from above, the width of the non-Bragg scattering
component decreases and the $q=0$ intensity increases.
Similarly, as $T \rightarrow T_N$ from below, the width decreases and the
$q=0$ intensity increases.  Such behavior is typical of an antiferromagnetic
phase transition.  To fit the data, we used the Lorentzian term in Eq.\ 1,
convoluted with the instrumental resolution.
The data for $|q|<0.008$~rlu were
eliminated from the fits to the Lorentzian term to avoid Bragg scattering.
The results of the fits yield $\kappa(T)$ and
the staggered susceptibility $\chi(T) = A/ {\kappa ^2}$.
The results for $\kappa$ are shown in
Fig.\ 3, along with the expected random-exchange critical
behavior\cite{bkj86,bfmspr98} as
indicated by the solid curves with $\nu = 0.69$ and
$\kappa_o ^+ / \kappa_o ^- =0.69$.  The overall amplitude of the solid curves is
adjusted to approximately follow the data.  A clear minimum
$\kappa \approx 0.017$ rlu is observed in the
fitted values near $T_N$, indicating significant rounding
due to a concentration gradient in the
crystal\cite{bkfj87}.  The gradient rounding is most likely the cause
of the deviations of the data from the fit away from the minimum as
well.  Nevertheless, the present data are
plausibly consistent with random-exchange critical behavior when the
significant rounding due to the concentration gradient is taken into
 account.

Results for the logarithm of $\chi$ vs.\ $T$ are shown in
Fig.\ 4.  The random-exchange behavior\cite{bkj86,bfmspr98},
with $\gamma = 1.31 $ and
${\chi_o} ^+{/\chi_o} ^- = 2.8$ and with the overall amplitude
adjusted to approximately fit the data, is shown as the solid curves.
The maximum in the data and the systematic deviations from the fit
are indications of a significant gradient in the concentration,
as we discussed with respect to Fig.\ 3.
Again the data are fairly consistent with a concentration-rounded
random-exchange transition to antiferromagnetic long-range order.

The Bragg intensity, obtained by subtracting the fitted Lorentzian
scattering intensity from the total $q=0$ scattering intensity,
is shown vs.\ $T$ in Fig.\ 5.  Once again, the data are fairly consistent
with a random-exchange\cite{rklhe88} transition ($\beta = 0.35$)
near $T=11K$ represented
by the solid curve.  The nonzero Bragg component above $T=11K$ is
probably attributable mainly to concentration gradient effects.
The precise shape of the Bragg scattering intensity vs.\ $T$
in Fig.\ 5 must not be taken too seriously, particularly
at low $T$, since it is known that severe extinction effects distort
the behavior by saturating the measured value\cite{by91}.  In addition, for
$T < 7$~K, the sample shows nonequilibrium effects since it was
quenched, as described above, and the magnitude of the Bragg scattering
component might well be smaller than if the sample were in equilibrium.
The large Bragg scattering component well below $T_N$, along with
the minimum in $\kappa$ and maximum in $\chi$ near $T_N$ strongly
indicate an antiferromagnetically ordered phase.

Previous magnetization studies\cite{mjmmr92,mrjmm95} indicate a de
Almeida-Thouless-like (AT)
curve in the $H-T$ phase diagram. The $H=0$ endpoint of this
boundary coincides reasonably well with the antiferromagnetic
phase transition observed with neutron scattering.

In conclusion, we have shown neutron scattering evidence that this
system, $Mn_{0.35}Zn_{0.65}F_2$, orders near $T=11$~K in a way consistent
with the REIM model.  In addition, significant relaxation
takes place for $T<7$~K.  This is consistent with previous magnetization
measurements and demonstrates that only part of the system orders with
long range order when the system is quenched to low temperatures.
This behavior is consistent with clusters coexisting with long
range order below $T_N$.  A similar glassy low temperature region has
been identified\cite{mkjhb91,mlcr90,jnm99} in the anisotropic system
$Fe_{0.31}Zn_{0.69}F_2$ using magnetization and dynamic susceptibility
measurements.  However, the broad line shapes
that indicate the glassy behavior were
not observed in $Fe_{0.31}Zn_{0.69}F_2$ with neutron scattering
techniques\cite{bmmkj91}.  It is interesting that neutron scattering measurements
at the percolation threshhold in $Mn_{0.25}Zn_{0.75}F_2$ did not
indicate any glassy behavior in contrast to $Fe_{0.25}Zn_{0.75}F_2$.
This should be investigated further.

This work has been supported
by DOE Grant No. DE-FG03-87ER45324
and by ORNL, which is managed by
Lockheed Martin Energy Research Corp. for the U.S. DOE
under contract number DE-AC05-96OR22464.
One of us (F.C.M.) also acknowledges the support of
CAPES, CNPq, FACEPE and FINEP (Brazilian agencies).

\begin{figure}[t]
\centerline{\hbox{
\psfig{figure=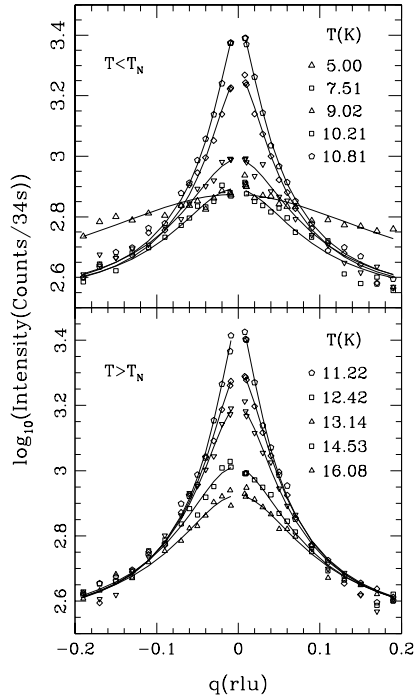,height=4.in}
}}
\caption{The logarithm of the neutron scattering intensity
vs.\ $q$ just above and below $T_N$ in zero field obtained
after quenching the sample to $T=5$~K and heating.}
\end{figure}

\begin{figure}[t]
\centerline{\hbox{
\psfig{figure=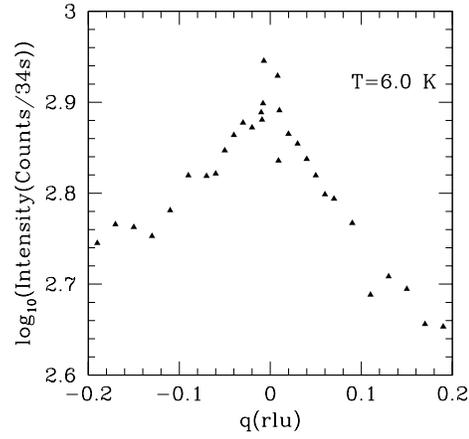,height=2.5in}
}}
\caption{The logarithm of the neutron scattering intensity
vs.\ $q$ for $T=6.00$~K in zero field obtained
after quenching the sample to $T=5$~K and heating.  The data were
taken in sequence in increasing $q$ with approximately $35$~seconds per
point.  The asymmetry indicates that on this time scale the system
is relaxing toward the behavior seen at higher $T$.}
\end{figure}

\begin{figure}[t]
\centerline{\hbox{
\psfig{figure=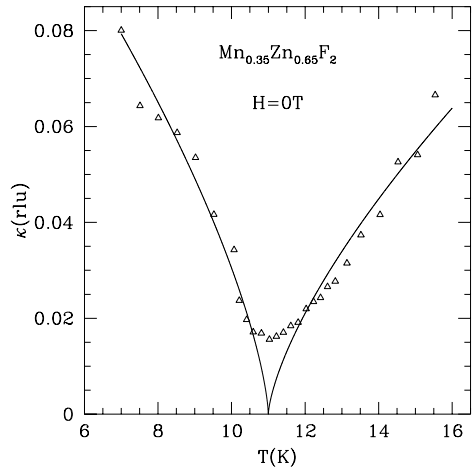,height=2.5in}
}}
\caption{$\kappa$ vs.\ $T$ near $T_N$.  The solid
curves represent the expected random-exchange critical behavior.}
\end{figure}

\begin{figure}[t]
\centerline{\hbox{
\psfig{figure=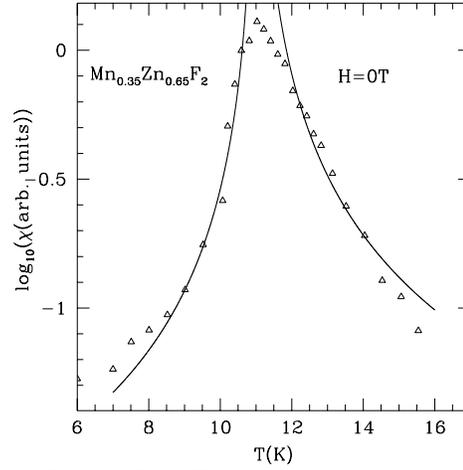,height=2.5in}
}}
\caption{The logarithm of $\chi$
vs.\ $T$ near $T_N$. The solid
curves represent the expected random-exchange critical behavior.}
\end{figure}

\begin{figure}[t]
\centerline{\hbox{
\psfig{figure=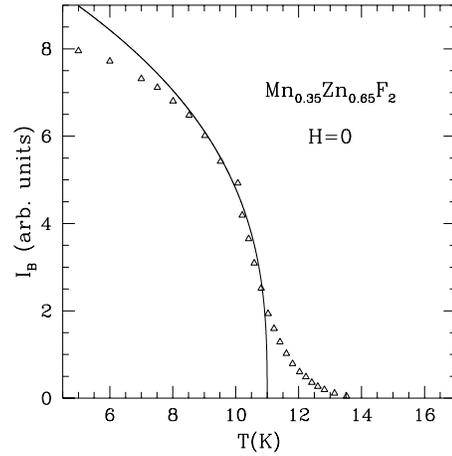,height=2.5in}
}}
\caption{The Bragg scattering intensity. The solid
curves represent the expected random-exchange critical behavior.
vs.\ $T$.  }
\end{figure}

\end{document}